\documentstyle[twoside]{article}

\catcode`\@=11
\long\def\@makefntext#1{
\protect\noindent \hbox to 3.2pt {\hskip-.9pt
$^{{\eightrm\@thefnmark}}$\hfil}#1\hfill}               

\def\@makefnmark{\hbox to 0pt{$^{\@thefnmark}$\hss}}    

\def\ps@myheadings{\let\@mkboth\@gobbletwo
\def\@oddhead{\hbox{}
\rightmark\hfil\eightrm\thepage}
\def\@oddfoot{}\def\@evenhead{\eightrm\thepage\hfil
\leftmark\hbox{}}\def\@evenfoot{}
\def\sectionmark##1{}\def\subsectionmark##1{}}

\oddsidemargin=\evensidemargin
\newcounter{sectionc}\newcounter{subsectionc}\newcounter{subsubsectionc}
\renewcommand{\section}[1] {\vspace{12pt}\addtocounter{sectionc}{1}
\setcounter{subsectionc}{0}\setcounter{subsubsectionc}{0}\noindent
        {\tenbf\thesectionc. #1}\par\vspace{5pt}}
\renewcommand{\subsection}[1] {\vspace{12pt}\addtocounter{subsectionc}{1}
      \setcounter{subsubsectionc}{0}\noindent
      {\bf\thesectionc.\thesubsectionc.{\kern1pt \bfit #1}}\par\vspace{5pt}}
\renewcommand{\subsubsection}[1]
      {\vspace{12pt}\addtocounter{subsubsectionc}{1}
      \noindent{\tenrm\thesectionc.\thesubsectionc.\thesubsubsectionc.
      {\kern1pt \tenit #1}}\par\vspace{5pt}}
\newcommand{\nonumsection}[1] {\vspace{12pt}\noindent{\tenbf #1}
        \par\vspace{5pt}}

\newcounter{appendixc}
\newcounter{subappendixc}[appendixc]
\newcounter{subsubappendixc}[subappendixc]
\renewcommand{\thesubappendixc}{\Alph{appendixc}.\arabic{subappendixc}}
\renewcommand{\thesubsubappendixc}
        {\Alph{appendixc}.\arabic{subappendixc}.\arabic{subsubappendixc}}

\renewcommand{\appendix}[1] {\vspace{12pt}
        \refstepcounter{appendixc}
        \setcounter{figure}{0}
        \setcounter{table}{0}
        \setcounter{lemma}{0}
        \setcounter{theorem}{0}
        \setcounter{corollary}{0}
        \setcounter{definition}{0}
        \setcounter{equation}{0}
        \renewcommand{\thefigure}{\Alph{appendixc}.\arabic{figure}}
        \renewcommand{\thetable}{\Alph{appendixc}.\arabic{table}}
        \renewcommand{\theappendixc}{\Alph{appendixc}}
        \renewcommand{\thelemma}{\Alph{appendixc}.\arabic{lemma}}
        \renewcommand{\thetheorem}{\Alph{appendixc}.\arabic{theorem}}
        \renewcommand{\thedefinition}{\Alph{appendixc}.\arabic{definition}}
        \renewcommand{\thecorollary}{\Alph{appendixc}.\arabic{corollary}}
        \renewcommand{\theequation}{\Alph{appendixc}.\arabic{equation}}
        \noindent{\tenbf Appendix \theappendixc #1}\par\vspace{5pt}}
\newcommand{\subappendix}[1] {\vspace{12pt}
        \refstepcounter{subappendixc}
        \noindent{\bf Appendix \thesubappendixc. {\kern1pt \bfit #1}}
        \par\vspace{5pt}}
\newcommand{\subsubappendix}[1] {\vspace{12pt}
        \refstepcounter{subsubappendixc}
        \noindent{\rm Appendix \thesubsubappendixc. {\kern1pt \tenit #1}}
        \par\vspace{5pt}}

\newcommand{\smalllineskip}{\baselineskip=10pt}

\def\eightcirc{
\begin{picture}(0,0)
\put(4.4,1.8){\circle{6.5}}
\end{picture}}
\def\eightcopyright{\eightcirc\kern2.7pt\hbox{\eightrm c}}

\def\abstracts#1#2#3{{
        \centering{\begin{minipage}{4.5in}\baselineskip=10pt\footnotesize
        \parindent=0pt #1\par
        \parindent=15pt #2\par
        \parindent=15pt #3
        \end{minipage}}\par}}

\renewenvironment{thebibliography}[1]
        {\frenchspacing
         \ninerm\baselineskip=11pt
         \begin{list}{\arabic{enumi}.}
        {\usecounter{enumi}\setlength{\parsep}{0pt}
         \setlength{\leftmargin 12.7pt}{\rightmargin 0pt} 
         \setlength{\itemsep}{0pt} \settowidth
        {\labelwidth}{#1.}\sloppy}}{\end{list}}

\newcounter{itemlistc}
\newcounter{romanlistc}
\newcounter{alphlistc}
\newcounter{arabiclistc}

\newcommand{\fcaption}[1]{
        \refstepcounter{figure}
        \setbox\@tempboxa = \hbox{\footnotesize Fig.~\thefigure. #1}
        \ifdim \wd\@tempboxa > 5in
           {\begin{center}
        \parbox{5in}{\footnotesize\smalllineskip Fig.~\thefigure. #1}
            \end{center}}
        \else
             {\begin{center}
             {\footnotesize Fig.~\thefigure. #1}
              \end{center}}
        \fi}

\newcommand{\tcaption}[1]{
        \refstepcounter{table}
        \setbox\@tempboxa = \hbox{\footnotesize Table~\thetable. #1}
        \ifdim \wd\@tempboxa > 5in
           {\begin{center}
        \parbox{5in}{\footnotesize\smalllineskip Table~\thetable. #1}
            \end{center}}
        \else
             {\begin{center}
             {\footnotesize Table~\thetable. #1}
              \end{center}}
        \fi}

\def\@citex[#1]#2{\if@filesw\immediate\write\@auxout
        {\string\citation{#2}}\fi
\def\@citea{}\@cite{\@for\@citeb:=#2\do
        {\@citea\def\@citea{,}\@ifundefined
        {b@\@citeb}{{\bf ?}\@warning
        {Citation `\@citeb' on page \thepage \space undefined}}
        {\csname b@\@citeb\endcsname}}}{#1}}

\newif\if@cghi
\def\cite{\@cghitrue\@ifnextchar [{\@tempswatrue
        \@citex}{\@tempswafalse\@citex[]}}
\def\citelow{\@cghifalse\@ifnextchar [{\@tempswatrue
        \@citex}{\@tempswafalse\@citex[]}}
\def\@cite#1#2{{$\null^{#1}$\if@tempswa\typeout
        {IJCGA warning: optional citation argument
        ignored: `#2'} \fi}}

\def\@refcitex[#1]#2{\if@filesw\immediate\write\@auxout
        {\string\citation{#2}}\fi
\def\@citea{}\@refcite{\@for\@citeb:=#2\do
        {\@citea\def\@citea{, }\@ifundefined
        {b@\@citeb}{{\bf ?}\@warning
        {Citation `\@citeb' on page \thepage \space undefined}}
        \hbox{\csname b@\@citeb\endcsname}}}{#1}}

\def\@refcite#1#2{{#1\if@tempswa\typeout
        {IJCGA warning: optional citation argument
        ignored: `#2'} \fi}}

\def\refcite{\@ifnextchar[{\@tempswatrue
        \@refcitex}{\@tempswafalse\@refcitex[]}}

\def\pmb#1{\setbox0=\hbox{#1}
        \kern-.025em\copy0\kern-\wd0
        \kern.05em\copy0\kern-\wd0
        \kern-.025em\raise.0433em\box0}

\def\fnt#1#2{\footnotetext{\kern-.3em
        {$^{\mbox{\scriptsize #1}}$}{#2}}}

\def\fpage#1{\begingroup
\voffset=.3in
\thispagestyle{empty}\begin{table}[b]\centerline{\footnotesize #1}
        \end{table}\endgroup}

\def\runninghead#1#2{\pagestyle{myheadings}
\markboth{{\protect\footnotesize\it{\quad #1}}\hfill}
{\hfill{\protect\footnotesize\it{#2\quad}}}}
\font\tenrm=cmr10
\font\tenit=cmti10
\font\tenbf=cmbx10
\font\bfit=cmbxti10 at 10pt
\font\ninerm=cmr9

\font\eightrm=cmr8

\def\qed{\hbox{${\vcenter{\vbox{                      
   \hrule height 0.4pt\hbox{\vrule width 0.4pt height 6pt
   \kern5pt\vrule width 0.4pt}\hrule height 0.4pt}}}$}}

\begin{document}

\runninghead{S. C. Tiwari
}
{Reality of time  $\ldots$}

\thispagestyle{empty}\setcounter{page}{184}
\vspace*{0.88truein}
\fpage{184}

\centerline{\bf REALITY OF TIME}
\vspace*{0.035truein}

\vspace*{0.37truein}
\centerline{\footnotesize S. C. Tiwari}

\centerline{\footnotesize \it
Institute of Natural Philosophy}
\baselineskip=10pt
\centerline{\footnotesize \it
c/o 1 Kusum Kutir, Mahamanapuri,
  Varanasi-221 005, India.}


\baselineskip 5mm

\vspace*{0.21truein}

\abstracts{
The meaning of instantaneous action at a distance is elucidated. It is
shown that the absence of a medium to transmit action (usually referred
to as AAD in the literature) and instantaneous action are not identical.
Since the term ``instantaneous" is incompatible with relativity and field
theory, a critique is presented on the concept of time in relativity. It is
argued that relativity does not deal with the nature of time. Physical
reality of the absolute time is envisaged, and instantaneous action is
proposed to be a natural consequence of it. Possibly gravity is such kind
of a force, however the electromagnetic force, though envisaged as direct
particle interaction (i.e. without intermediary fields) can not be
instantaneous.}{}{}


\bigskip

$$$$

\section{Introduction}

Sense impressions or empirical observations seem to suggest a dichotomy of
the physical world into isolated independently existing objects and a
physical mechanism to relate observed mutual influences between them. In
modern perspective one finds expression of this dichotomy in terms of
elementary particles and their interactions(believed to be the four
fundamental forces of nature). Philosophical urge to seek connectedness
with the separation, and wholeness within the parts have led to the
conception of physical reality with intrinsic duality. Primarily due to
this, both action-at-a-distance (AAD) and field theory possess a sort of
indistinguishable character as far as the verifiable experimental
consequences are concerned. The main aim of this article is to elucidate
the proposition that the nature of time might play a fundamental role to
throw light on the meaning of instantaneous AAD. To delineate the basic
problems a brief commentary on AAD is presented in the next section. The
concept of time is discussed in section 3. In the final section the idea
of the absolute time is presented, and its implications on AAD versus
field theory are outlined.

\section{Review of Action-At-A-Distance}

The idea of AAD is commonly attributed to Newton, however, Cajori in the
appendix to Newton's {\it Principia} [1] argues that the doctrine of AAD
belongs to Cotes, not to Newton. He quotes Maxwell to support his
claim,and refers to a letter from Newton:
\begin{quotation}
\noindent
``{\sl That gravity should be innate, inherent
and essential to matter, so that one body may act upon another at a
distance through a vacuum, without the mediation of anything else, by and
through which their action and force may be conveyed from one to another,
is to me so great an absurdity, that I believe no man,who has in
philosophical matters a competent faculty of thinking,can ever fall into
it. Gravity must be caused by an agent according to certain laws,but
whether this agent is material or immaterial, I have left to the
consideration of my readers}."
\end{quotation}

The transition from `absence of a medium'
to `instantaneous action' is not straightforward, in fact the discoveries
of force laws in electricity and magnetism, and analogy with Newton's
gravitational force law led to a subtle change in the
interpretation. Newtonian interpretation suggests that a medium to
transmit force is necessary, but it can happen instantaneously. Maxwell
field equations, on the other hand give rise to wave equations such that
the fields have finite velocity of propagation.  For an electrostatic
field of a charged body an additional feature emerges: the longitudinal
waves. To see this, the electrostatic potential $\Phi({\bf r})$
is expanded in a Fourier integral, and the Poisson equation is used to
obtain the following relation for a point charge
\begin{equation}
\Phi_{\bf k}=\frac{4\pi e}{k^2},
\end{equation}
here $\Phi_{\bf k}$ is the
Fourier component of a zero frequency wave with wave vector ${\bf
k}$. Fourier components of electric field vector can be calculated using
equation (1) to give
\begin{equation}
{\bf E}_{\bf k}=-i\frac{4\pi e {\bf k}}{k^2}.
\end{equation}
Since ${\bf E}_{\bf k}$ is parallel to ${\bf k}$, the Coulomb
field has longitudinal waves, and the time independence is interpreted to
imply instantaneous AAD. Generalizing the Poisson equation to
in-homogeneous wave equation the action propagating with the velocity of
light is obtained. One can formulate a theory without introducing the
fields using the interaction action integral:
\begin{equation}
{\cal L}_I=\frac{1}{2}\sum\limits_{i,j=1(i\neq j)}^n e_i e_j \int
\frac{dz_i^\mu}{d\lambda_i}\frac{dz_\mu^j}{d\lambda_j}\delta
\left[(z_i-z_j)^2\right]d\lambda_id\lambda_j.
\end{equation}
This type of theory
is called AAD because no mediating agency is required to carry the
interaction [2]. Here $\lambda_i$ is a parameter used to label the
worldline $z_i^\mu$ of $i$-th charged particle $e_i$ of the $n$-particle
system. Delta-function in (3) ensures that for ${\cal L}_I$ to be
non-vanishing the ``worldlines" are connected by light signal, and self
interaction is absent as $i\neq j$. The statement in [3] ``{\it strictly,
the phrase `action at a distance' should be changed to `action at no
distance'}" shows that such theories are not genuine AAD.  Moreover,
direct particle fields introduced in this approach, and the perfect
absorption by the rest of the universe postulated by Wheeler and Feynmann
[4] though explain the existence of radiation overcoming the difficulties
faced by the Schwarzschild-Tetrode-Fokker theory, the resulting scheme is
more akin to field theory than the AAD. Ironically the delta-functions in
field theories serve the purpose of bringing in particulate aspect as
field singularities while in this version of the AAD they incorporate
fields. The simplest way to recognize this fact is to affect the variation
on ${\cal L}_I$, and define the direct potential as
\begin{equation}
A_i^\mu(z_k)=e_i\int\limits_{-\infty}^{+\infty}\delta
\left[(z_k-z_i)^2\right]\frac{dz_i^\mu}{d\lambda_i}d\lambda_i.
\end{equation}
These
potentials satisfy the inhomogeneous wave equation, and depend on the
sources more like defining relations without independent degrees of
freedom. The details can be seen in [2]. The application of this approach
to derive the equation of motion of electron has been extensively
discussed in the literature. An important critique by Havas [5] deserves
attention comparing Dirac's theory with this theory arriving at the
conclusion that experimentally the two theories are ``indistinguishable".

An important problem which has a bearing
on this discussion, the essence of which was perceived by Planck studying
the matter-radiation interaction arises in the quantum
electrodynamics. Einstein's light quantum hypothesis and field theory
seemed irreconcilable to him, and led to his second and third
theories [6]. In 1911 he proposed that while emission of radiation occurs
discontinuously in quanta, absorption is a continuous process. In this
second theory, a remarkable result followed, namely the appearance of
zero-point energy of the oscillator. In 1914 he revised his ideas and
proposed his third theory: the emission and absorption are both
continuous, the discrete energy exchange takes place only during
collisions between the oscillators and free particles. In the language of
quantum electrodynamics, the interaction is described in terms of the
exchange of photons i.e. the carriers of the interaction. Following Planck
we may ask if the  photon acquires continuous field property once emitted
from the charged body. Or is it possible to formulate a complete particle
theory totally dispensing with the continuous electromagnetic fields, and
employing only photons?

Present brief commentary shows that the role of time, the nature of sources
and the understanding of continuum are fundamental issues to be addressed
for making further decisive progress in resolving the debate on the
mechanism of interaction(s), and that one must be careful as regards to
the version of AAD one is referring to.

\section{The Concept of Time: A Critique}

Relativist's world-view is claimed to have revolutionized our thinking on
space and time, radically altering the classical Newtonian system. Is it
true? To answer this question, first I will base my discussion on
Einstein's own exposition of his theory of relativity[7]. Special theory
of relativity (STR) asserts the following: (1) The laws of nature are in
concordance for all inertial frames, (2) the hypothesis of the absolute
character of time is discarded, (3) simultaneity has no objective meaning,
and (4) four dimensional space-time has absolute character, and has
physical reality.  Einstein does not define motion, and assuming uniform
motion postulates inertial frames ($K$). Absolute time, according to him
has the meaning: ``{\it the time of an event in $K'$ is the same as the
time in K}".  This absolute time makes physical sense if instantaneous
signals exist, and the state of motion of a clock has no effect on its
rate. Since no such signals exist, a scheme for time measurement is
suggested postulating the constancy of the velocity of light in vacuum. In
a frame $K$, clocks $U_n$ relatively at rest are placed at all points
of space. A light signal from one of the clocks $U_k$ sent at the instant
$t_k$ travels in vacuum to $U_l$ placed at a distance $d_{kl}$ which is
now set to indicate the time
\begin{equation}
t_l=t_k+\frac{d_{kl}}{c}.
\end{equation}

In all inertial systems the clocks are regulated in this manner.
Simultaneity becomes relative, and time specification depends on the space
of reference. The constancy of the velocity of light in all inertial
frames leads to a geometrical invariant 4-dimensional length
\begin{equation}
ds^2=dr^2-c^2dt^2=0.
\end{equation}
The step from (6) to arbitrary non-zero $ds^2$ being invariant invokes the
assumption of the directional independence in space. In analogy with the
Euclidean geometry, the world-geometry in STR is postulated to be
4-dimensional space-time continuum.  In contrast to Einstein, Newton is
careful to distinguish two kinds of space, time and motion: absolute and
relative. Absolute (relative) motion is the translation of a body from one
absolute (relative) place into another. Absolute time, by its own nature
flows equably without relation to anything external. Relative or common
time is some sensible and external measure of duration by means of motion
[1]. Einstein is concerned with this relative time, and the process of its
measurement. What is the meaning of the instant $t_k$ recorded by the
clock $U_k$? Obviously it is implicit assumption that there is something
called time, and that there exists earlier and later ordering of time.

Criticisms/analyses by other authors have been presented in some of my
earlier papers [8], however, a brief discussion may be pertinent at this
point. Reichenbach using light and matter axioms offers an axiomatization
of STR with space, time and motion as presupposed concepts [9]. He says
that modern epistemology has shown that Kant's intuitive time is
untenable, and attempts to define time at a point P without invoking the
direct perception of time flow. First a signal is defined as a physical
process which can be marked at a real point $P$, and can be transmitted to
another point $P'$ where the mark can be recognized. A signal marked at
$P$ is sent to some other point $P'$, again marked at $P'$ returns to $P$.
The detection of two marks ascertains the temporal sequence at $P$. It is
logically unsatisfactory to consider another distant point $P'$ in order
to define the time order at $P$. Not only this, Reichenbach ignores the
interaction duration for the marking process, and the memory and logical
decision making capability of the detection device.

Bunge's aim is to free STR from misinterpretations [10]. Time, Euclidean
space and inertial frames belong to proto-physics, and the role of
electromagnetic fields is stressed by him. Newtonian absolute time is
criticized as being metaphorical and hanging in the midst of
nothingness. `Flows equably' implies constant rate which in turn involves
the notion of time, and hence there is a circularity in this
definition. However, Bunge's analysis does not provide any new insight into
the meaning of time, and temporal order as well as protophysics remain
unexplained.

Thus relativity is founded on motion, inertial system and relative time of
Newton with a new measurement convention using light signal, see also
Dingle's interpretation[11]. It does not deal with the absolute time and
its nature. I have argued earlier [8] that it is not merely a question of
philosophical taste or pure reasoning, the life-time of an unstable
particle changing with velocity poses a paradox if relativistic time
dilation is identified with this. The decay being an irreversible
process, life-time corresponds to a unidirectional lapse of time;on the
other hand time dilation is based on time inversion symmetrical
kinematics.  Therefore, the question of the physical reality of time is
not an empty issue; we address this problem in the next section.

\section{Absolute Time: Physical Reality}

The basic idea put forward in late 1970s [12] is very simple: time is the
primeval cause of the manifest universe, and time flow is regulated by a
Cosmic principle. To make physical sense out of it, and to relate it with
the contemporary understanding and the language of discourse is, however
not easy. We postulate an inert and unobservable source space which is
continuously being changed to observable active space(or manifest universe)
by the action of time. The possible active states are proposed to be those
of rotation and expansion. An alternative definition of source space is the
rest of the space other than the universe. The universe spins with
reference to the source space and expands into it. The cause of rotation
and expansion is the absolute time, the duration of which is determined by
cosmic motion.  Universality of the absolute time flow entails space
continuum for the whole of the universe. The rotation determines
periodicity of the cosmic clock, and expansion makes the time to be
inherently unidirectional. Time reversal is non-existent, and space-time
continuum has no meaning for this absolute time. Though the spatial changes
within the universe are correlated with the absolute time frame as far as
earlier and later ordering is concerned, one is free to setup convenient
definition and measurement setup for relative or common time to describe
such spatial relations.

In a somewhat abstract construction following[8]
the absolute and common time can be explained noting that any measurement
in physics ultimately reduces to a counting of spatial relations in terms
of a reference standard unit, $D$. For unchanging spatial relations between
any two points in the space (a rigid structure) one can use a standard
scale, but this method has to be modified once changing spatial relations
are allowed. A reference unit $C$ possessing intrinsic uniform spatial
relation, if it exists may replace the scale for calibrating changing
spatial relations. Let us imagine a space $S$ within which occur continuous
flow of space points thereby leading to changing spatial relations. Note
that this description endows the space $S$ an active role i.e. moving
points. One can classify sets of uniform changing relations, $I$, and admit
the possibility of reversible changes. To reduce the complexity of the
description an order parameter, $t$ may be introduced, and calibrated using
the device $C$. Let us now suppose some `external' change on the whole of
the space $S$ such that this regulates the changes within $S$ i.e.  there
is a harmony between the two, and this change, denoted by $T$ takes place
uniformly and unidirectionally without relation to anything 'inside'
$S$. Translating this description to physical situation, $S$ is the
universe, $t$ is common time, $I$ is the inertial frames, $C$ is the light
signal, and $T$ is the absolute time. The parameter $t$ merely represents
a convenient mathematical definition to coordinate the changes in spatial
relations, and admits the transformation $t\rightarrow -t$, but it is
unphysical.  The parameter $t$ acquires physical meaning only when it is
related with unidirectional flow $T$.

Inadvertently identity of mathematical parameter and physical common time
has been assumed since the time of Newton resulting into unexplainable
possible physical situations. In STR the light cone and its future and past
division illustrates one of the problems: equation (6) is valid for
mathematical time $t$, but to relate it with T one invokes causality and
impossibility of any material particle attaining or exceeding the
velocity of light.  In general relativity the things are not so simple
since the metric in
\begin{equation}
ds^2={\sl g}_{\mu\nu}dx^\mu dx^\nu
\end{equation}
will change the light cone from one space-time point to another. Usually
one defines locally special relativistic time in the GR, but near massive
bodies this procedure is not applicable. Singularities as space-time edge
appear to be distinct at the beginning (e.g. big-bang creation) and end
(collapse of a massive object to a black hole), see [13]. Does it indicate
arrow of time? The question itself becomes meaningless as shown by Godel's
solution [14]. He points out that relativity of simultaneity implies that
lapse of time has no objective meaning [15]. In cosmological models based
on GR which appear to be consistent with the observations, a time common
to the whole universe can be defined. Jeans took it as supporting the idea
of an absolute lapse of time, but Godel argues that there do exist
cosmological solutions which allow one to travel in the past, present and
future. The assumed metric is a solution of the field equation
\begin{equation}
R_{\mu\nu}-\frac{1}{2}{\sl g}_{\mu\nu}R=8\pi k\varrho u_\mu u_\nu +
\lambda{\sl g}_{\mu\nu}.
\end{equation}

Besides the matter energy-momentum tensor, it has the cosmological constant
term first introduced by Einstein for static universe, but with opposite
sign
\begin{equation}
\lambda=-\frac{R}{2}(=-4\pi k\varrho).
\end{equation}

Using time-like and null vectors a reasonable definition of time direction
is introduced, but it is found that no time coordinate at each space-time
point exists which increases as one moves in a positive time-like
direction. Einstein in his reply [15] very clearly recognizes the
difficulty of incorporating time ordering in relativity, but insists on
the necessity of a signal to define time. Regarding Godel's solution he
asks whether these are acceptable on physical grounds. Unfortunately there
is no physical or mathematical criterion to choose a field equation
like (8) amongst many possible field equations;and even if one settles down
to a definite field equation there exist arbitrary large number of
solutions of the field equation. In view of the idea of the absolute time
$T$ we conclude that time in GR is not physical time.  Newtonian absolute
time is regarded as a metaphysical concept by Cajori [1] but $T$ discussed
here has physical reality. Profound consequences follow from this idea as
shown in a recent monograph [16]. Here we limit our discussion to its
implications in the context of AAD. By its very nature universality of
time $T$ in the whole of the universe implies space continuum, and
instantaneous action. Thus the space could be viewed as a medium in
Newtonian sense, and if the meaning of AAD is understood to be the absence
of a medium then this action is instantaneous but not AAD. In what form
does this action manifest in the universe? In [12] it was suggested that
gravitation as an apparent force originates due to inhomogeneous
distribution of spatial structures and the expansion of the
universe. Therefore, identifying gravitation with this action, we arrive
at the conclusion that gravitation is an instantaneous action.  Let us now
consider the electromagnetic force. The first significant out-come of our
space-time analysis is that the nature of source i.e. electric charge can
be understood in mechanical terms. This becomes possible with the belief
that cosmic process is replicated at sub-atomic level, and the observation
that the electric charge occurs as $e^2$ once the field quanti- ties are
expressed in geometrical units(for example electric field in the unit of
length$^{-2}$). That large scale structure of universe has some deep
connection with the  atomic structure is a belief shared by many
physicists, notably by Eddington as pointed out by Infeld in [15].
Einstein rejected it but recent unified theories also seem to indicate
this connection. A space-time model is envisaged such that
``spatio-temporal" bounded structures posses spinning motion and
translation in space (corresponding to rotation and expansion of the
universe respectively). Noting that the dimension of $e^2/c$ is that of
angular momentum, we can formulate a direct particle interaction theory
for electromagnetism. The localized space-time structures approximated as
isolated objects are defined to be the particles, e.g. electron and
photon. Electron-electron or electron-photon interaction is a direct
particle contact interaction(collision) not action-at-a-distance, and
there is no self field associated with them. External electric or magnetic
fields represent a kind of photon fluid. Two electrons immersed in such a
photon fluid placed a distance apart transmit the disturbance caused by
their spinning motion through the fluid, and influence each other. This
interaction is therefore not instantaneous.  Since this approach is
radically different than the standard electromagnetism, there are many
open problems in it though some progress has been made and we refer the
reader to[16]. I believe alternative paradigm for fundamental physics
propounded here deserves attention to resolve the debatable issues such as
instantaneous AAD versus field theory.

\nonumsection{Acknowledgments}

I am gratefull to Dr. R. Smirnov-Rueda for
inviting me to write this essay which proved to be a rewarding experience.
Library facility at  the Banaras Hindu University is acknowledged.

\nonumsection{References}

\end{document}